\documentclass{bulsao}
\usepackage{graphicx}
\usepackage{latexsym}
\usepackage{longtable}

\begin{document}

\title{Speckle Interferometry of Metal-Poor Stars in the Solar Neighborhood.\,I}

\author{D.~A.~Rastegaev, Yu.~Yu.~Balega, E.~V.~Malogolovets}

\institute{Special Astrophysical Observatory, RAS, Nizhnii Arkhyz,
Karachai-Cherkessian Republic, 357147 Russia}

\offprints{D.~A.~Rastegaev, \email{leda@sao.ru}}

\date{received: June 15, 2007/revised: June 28, 2007}

\titlerunning{Speckle Interferometry of Metal-Poor Stars ...}
\authorrunning{Rastegaev et al.}

\abstract{
We report the results of speckle-interferometric observations of
109 high proper-motion metal-poor stars made with the 6-m
telescope of the Special Astrophysical Observatory of the Russian
Academy of Sciences. We resolve eight objects --- G102-20,
G191-55, BD+19$^\circ$~1185A, G89-14, G87-45, G87-47, G111-38,
and G114-25 --- into individual components and we are the first
to astrometrically resolve seven of these stars. New resolved
systems included two triple (G111-38, G87-47) and one quadruple
(G89-14) star. The ratio of
single-to-binary-to-triple-to-quadruple systems among the stars
of our sample is equal to 71:28:6:1.
}

\maketitle

\section{INTRODUCTION}
Stars of the halo and thick disk of our Galaxy are old metal-poor
objects with large spatial velocities
(\cite{norris:Rastegaev_n}; \cite{majewski:Rastegaev_n}). The studies of
these stars can be used to impose constraints on the physical
conditions during the early stages of the formation of our Galaxy.
Binary and multiple stars are the best candidate objects
to be used for studying the process of star formation at the time of
formation of our Galaxy, because they bear more information about this
process compared to single stars. This information is coded both
in the orbital parameters (eccentricity, semi-major axis) and
physical parameters of the components (component luminosities and
the mass function).

The authors of early papers dedicated to the study of the
multiplicity of the old population of the Galaxy
(\cite{abt_levi:Rastegaev_n}; \cite{abt_willmarth:Rastegaev_n}) concluded
that the fraction of binary and multiple systems among these
objects is very low compared to the corresponding fractions for
younger stars of the Galactic disk, which are richer in heavy
elements. However, the picture has changed in the last two decades.
 In  series of papers opened by
(\cite{paperI:Rastegaev_n}) is shown that the binary-to-single star ratio
for halo and thick-disk stars is comparable to the corresponding
ratio for the overwhelming majority of stars in the solar
neighborhood. Such studies are based on the analysis of stellar
spectra (\cite{paperXV:Rastegaev_n}; \cite{paperXVI:Rastegaev_n}) using the
data on visual binaries and common proper motion pairs
(\cite{allen:Rastegaev_n}; \cite{osorio:Rastegaev_n}).

We still have insufficient data about the multiple  old
stars in the solar neighborhood with orbital semi-major axes in the
interval from  $\sim$1 to $\sim$100\,AU observable with adaptive
optics and speckle-interfero\-metric facilities. We point out the
paper by Zinnecker et al. (\cite{zinnecker:Rastegaev_n}) who report the results of
observations of population~II stars using the techniques of
speckle interferometry, adaptive optics, and direct imaging. To
expand the database on such objects and determine the properties
of the components of multiple systems, we began
speckle-interferometric observations of metal-poor objects with
large proper motions  located in the close vicinity of the Sun.
In this paper we report the results of observations of 109 halo
and thick-disk stars made during the period from April through
December, 2006.

The paper has the following layout: Section 2 describes the
sample of stars studied; Section~3 analyzes the methods of
observations and reduction of the data obtained; Sections 4 and 5
list the results of observations and additional information about
the resolved stars, respectively; Section~6 discusses the
multplicity of the stars studied, and the last section gives the
conclusions.

\vspace*{0.3cm}
\section{THE SAMPLE}
We selected our program stars from the CLLA catalog
(\cite{clla:Rastegaev_n}). This catalog is actually a sample of A-
to early K-type dwarfs from the  \textit{Lowell Proper Motion
Catalog} (\cite{lpmc1:Rastegaev_n}; \cite{lpmc2:Rastegaev_n}), which
contains mostly Northern-Hemisphere stars brighter than 16
magnitude with proper motions exceeding $0.26\ ''/$year.

We selected a total of 223 stars from the CLLA catalog based on
the following three criteria:\\
\vspace{1cm}
~~~~ 1.~$\mathrm{[m/H]}<-1$;

2.~$\delta>-10^\circ$;

3. $\mathrm{m_V}<12$.

\begin{figure}
\includegraphics[width=8.5cm]{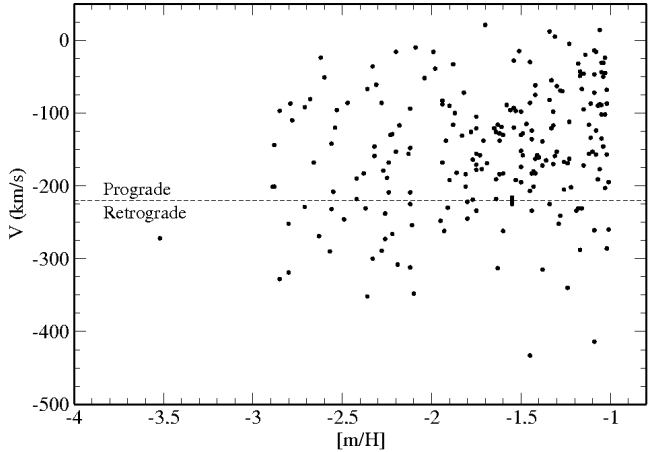}
\caption{Distribution of the stars of
the sample by metallicity [m/H] and V component of spatial velocity.
The dashed line separates stars moving in prograde and retrograde
orbits.} \label{m_H_vs_V:Rastegaev_n}
\end{figure}

\begin{figure}
\includegraphics[width=8.5cm]{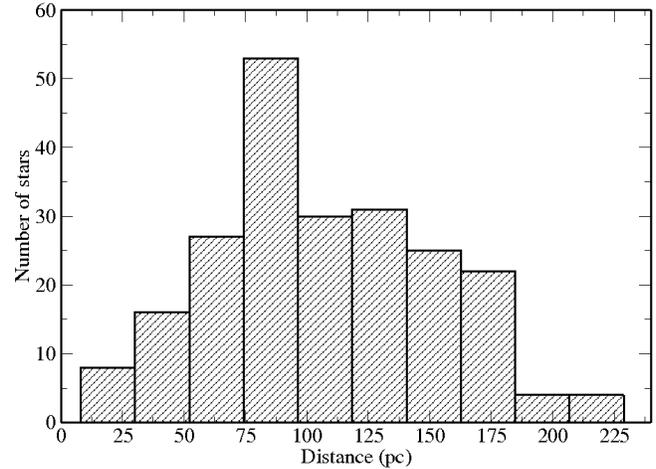}
\caption{Distribution of heliocentric
distances of the stars of the sample (we adopt the distances from
the  CLLA catalog).} \label{Distances:Rastegaev_n}
\end{figure}

\begin{figure}
\includegraphics[width=8.5cm]{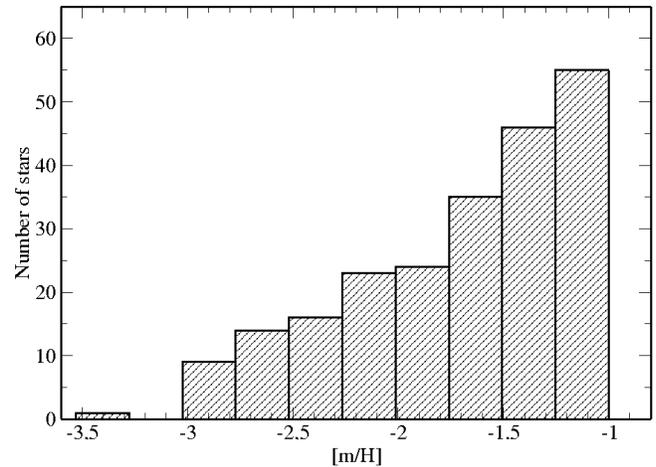}
\caption{Distribution of the metallicities
of the stars of the sample (we adopt the metallicities from the
CLLA catalog).} \label{metal:Rastegaev_n}
\end{figure}

\begin{figure}
\includegraphics[width=8.5cm]{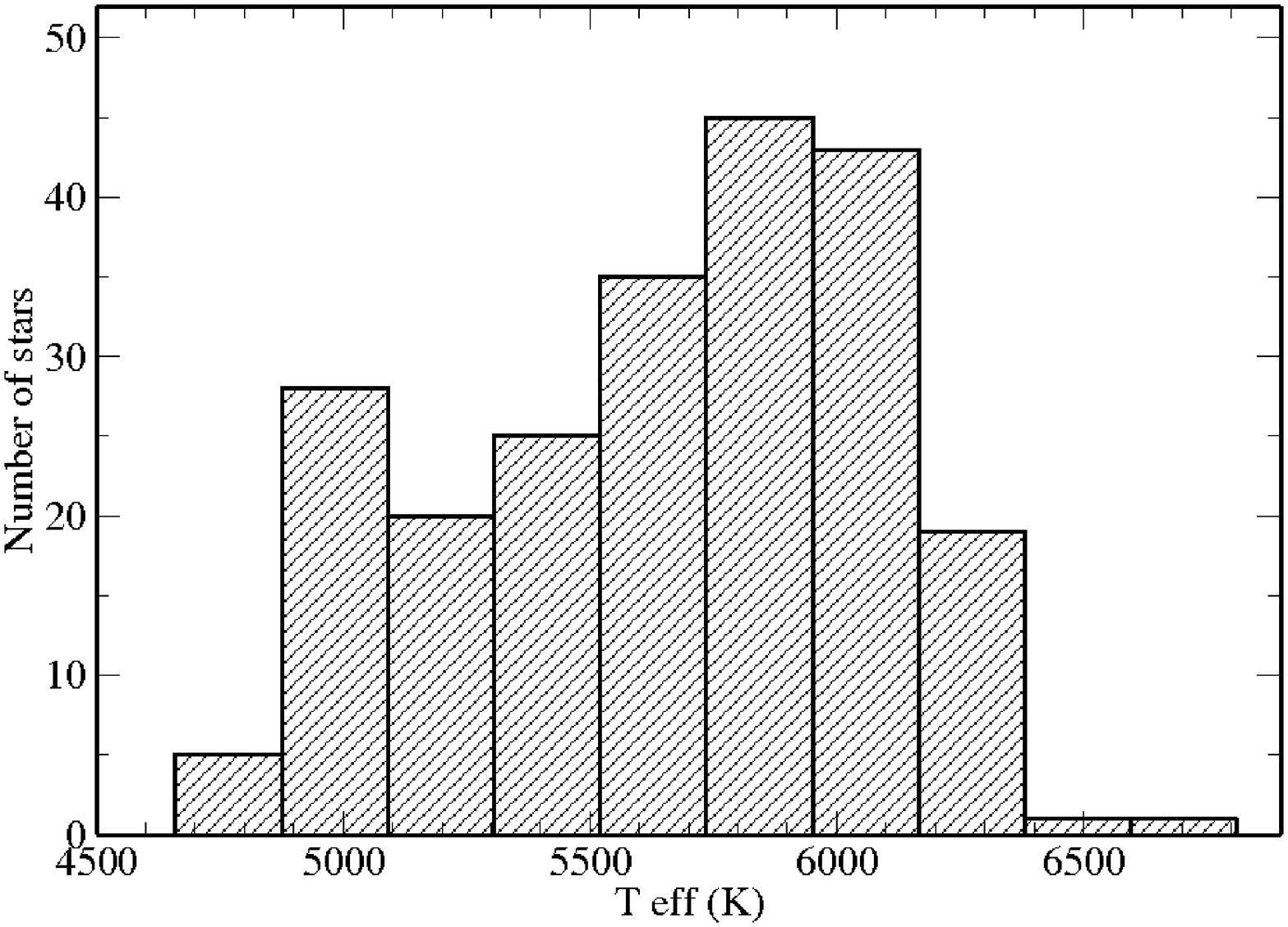}
\caption{Distribution of the temperatures of
the stars of the sample (we adopt the temperatures from the CLLA
catalog ).} \label{Teff:Rastegaev_n}
\end{figure}

We use the data of the CLLA catalog to make several figures
illustrating some of the main characteristics of our sample.
Figure~\ref{m_H_vs_V:Rastegaev_n} shows the distribution of our
selected stars on the  ``metallicity -- V-component of the spatial velocity'' plane,
which demonstrates that our stars belong to
different components of the Galaxy. The left-hand part of the
diagram is occupied by halo objects
--- metal-poor stars with high velocity dispersion. The upper right corner
is populated by stars belonging to the metal-weak tail of the
thick disk. About 20\% of all stars (45 objects) move in
retrograde orbits. Figure~\ref{Distances:Rastegaev_n} shows the
distribution of the heliocentric distances of the stars of our
sample. We used the photometric distances from the  CLLA catalog.
 Arifyanto et al.
(\cite{arifyanto:Rastegaev_n}) compared the photometric parallaxes
with the corresponding trigonometric parallaxes measured by
\emph{Hipparcos} (\cite{esa:Rastegaev_n}) for stars of the catalog
considered and showed that there is a small discrepancy between
the heliocentric distances determined using different methods.
However, we did not correct the photometric distances in any way.
Fig.~\ref{metal:Rastegaev_n} shows the distribution of the
metallicities of the stars studied. The $\mathrm{[m/H]}<-3$
metallicity interval is represented by only one star, G64-12,
with $\mathrm{[m/H]}=-3.52$. Half of the stars studied have
metallicities in the interval $\mathrm{[m/H]}=\left[-1.58;
-1\right)$. As is evident from the distribution of stellar
temperatures (Fig.~\ref{Teff:Rastegaev_n}), these are  F-, G-,
and K-type stars.

\section{OBSERVATIONS AND DATA REDUCTION}
We performed speckle-interferometry of 109 stars of the sample
with the 6-m telescope of the Special Astrophysical Observatory
of the Russian Academy of Sciences (SAO RAS): in April (one star), May
(five objects), June (six objects), and December, 2006 (97
objects). Before December 2006 observations were made using
a facility described by Maksimov et
al.~(\cite{maksimov03:Rastegaev_n}). Its detector consists of a fast
1280$\times$1024 Sony ICX085 CCD combined with
a three-stage image intensifier  with electrostatic focusing. In December
we used a new facility based on an EMCCD (a CCD with internal
electron gain) with higher quantum efficiency and better linearity.
Both facilities are capable of detecting objects with component
magnitude differences up to 4$^m$.

We recorded speckle interferograms in filters with the parameters
of $545/30$~nm (the first and the second numbers give the central
wavelength and half bandwidth of the filter, respectively),
$550/20$, $800/110$, and $800/100$~nm with exposures ranging from
five to 20 milliseconds. In December 2006 we obtained 500 short-exposure images for each of
most of the objects.   For each of
the remaining objects 2000 exposures are accumulated.

A description of the technique used to determine the relative positions
and component magnitude differences from the power spectra of
speckle interferograms averaged over the set can be found in \cite{balega02:Rastegaev_n}. The diffraction limit of the resolution
is equal to $0.022''$ in the $545/30$ and $550/20$~nm filters and
$0.033''$ in the $800/110$ and $800/100$ filters. The measured
position angles and angular separations $\rho$ are accurate
to 0.3--2.8$^{\circ}$ and 1 to 5~mas, respectively. The errors of measured
$\theta$ and $\rho$ depend on a number of parameters: component
separation, magnitude difference, and seeing $\beta$. The accuracy of the
component magnitude difference determined from the power spectrum
is also a function of the same parameters. It usually varies from 0.05$^m$
to 0.2$^m$ for $m_{V}=8$ -- $10$ objects.

\section{RESULTS OF OBSERVATIONS}
The main results of this work are listed in Tables~\ref{tab1:Rastegaev_n}
and ~\ref{unresolved:Rastegaev_n}.

We resolved into components the following eight objects:
G102-20, G191-55, BD+19$^\circ$~1185A, G89-14, G87-45, G87-47,
G111-38, and G114-25. We astrometrically resolved seven among these eight
objects for the first time and found new components in five objects:
G191-55, G89-14, G87-47, G111-38, and G114-25. We were the first
to astrometrically resolve the well-known spectroscopic binary \mbox{G102-20}
(\cite{paperXVI:Rastegaev_n}). We are also the first to establish the
astrometric binarity of the  G87-45 system whose spectrum exhibits
signatures of three components (\cite{paperXVI:Rastegaev_n}). The
astrometric binarity of the resolved object BD+19$^\circ$~1185A was
earlier discovered by \emph{Hipparcos} (\cite{esa:Rastegaev_n}).

Table~\ref{tab2:Rastegaev_n} gives some data on resolved systems.
The last column of this table summarizes the results of all available
astrometric and spectroscopic observations of the objects studied
including the speckle-interferometric observations performed with
the 6-m telescope of SAO RAS.
\section{ADDITIONAL DATA ON RESOLVED STARS}
In this section we summarize additional data on the resolved stars.
For some objects we list two distances determined from photometric
(\cite{clla:Rastegaev_n}) and trigonometric (\cite{esa:Rastegaev_n})
parallaxes. It is evident that the distance determined using the former
method is underestimated, because it does not take into account
the luminosity of the additional component. On the other hand, the
additional component also introduces certain error in the measured
trigonometric parallaxes, especially for short-period systems.

\vspace*{0.3cm}
{\bf G102-20} ($05^h40^m09\fs7$ $+12^\circ10' 41''$;  HIP~26676).
This known SB1-type spectroscopic binary with a 26-year period (\cite{paperXVI:Rastegaev_n})
and a heliocentric distance of \mbox{$\approx70$~pc} (\cite{esa:Rastegaev_n}) was resolved
speckle-interferometrically for the first time.

\vspace*{0.3cm}
{\bf G191-55} ($05^h57^m28\fs6$ $+58^\circ 40' 49''$; BD+58$^\circ$~876; TYC~3762-1904-1).
It is an F8-type binary (\cite{simbad:Rastegaev_n}) located
at a heliocentric distance of $\approx93$~pc (\cite{clla:Rastegaev_n}).

\vspace*{0.3cm}
{\bf BD+19$^\circ$\,1185A}~~~~ ($06^h03^m14\fs9$~~~~~~ $+19^\circ21' 39''$; HIP~28671).
It is a G0V-type object (\cite{simbad:Rastegaev_n}). The binary nature of this star
with $\rho=223$ mas was discovered by \emph{Hipparcos}. Its heliocentric
distance is estimated at $d\approx42$
(\cite{clla:Rastegaev_n}) and $d\approx66$~pc (\cite{esa:Rastegaev_n}).
This is a triple system given the presence of a distant companion
($\rho\approx7''$) BD+19$^\circ$\,1185B.

\vspace*{0.3cm}
{\bf G89-14} ($07^h22^m31\fs5$ $+08^\circ49' 13''$; HIP~35756; WDS~07224+0854)
is a new quadruple system. We resolved it as a pair with a separation
of $0.99''$. Allen et al.~(\cite{allen:Rastegaev_n}) provide evidence
suggesting that the system may contain  a physically bound component at a distance
of  $34''$. At the same time, G89-14 is an SB1-type spectroscopic binary
with a period of 190 days (\cite{paperXVI:Rastegaev_n}). The star is at a
heliocentric distance of  $\approx94$~pc (\cite{esa:Rastegaev_n}).

\vspace*{0.3cm}
{\bf G87-45} ($07^h32^m58\fs7$ $+31^\circ07' 00''$; TYC~2453-763-1). It is
a G2-type star (\cite{simbad:Rastegaev_n}) known as an SB2-type
spectroscopic binary with a period of 51 days (\cite{paperVI:Rastegaev_n}).
The spectrum of this star exhibits signs of a third component
(\cite{paperXVI:Rastegaev_n}), which we must have resolved. The distance
to the star is  $d\approx123$~pc (\cite{clla:Rastegaev_n}).

\vspace*{0.3cm}
{\bf G87-47} ($07^h35^m34\fs1$ $+35^\circ57' 11''$; HIP~36936).
A new triple system whose distance is estimated at $d\approx62$
(\cite{clla:Rastegaev_n}) and $d\approx100$~pc (\cite{esa:Rastegaev_n}).
The system is known as an SB1-type spectroscopic
binary with a period of 13 days. The Hipparcos catalog lists it as
an object with stochastic astrometric solution. We found a third
component. For this object we could determine the position of the
secondary only with an uncertainty of $\pm180^\circ$.

\begin{table*}[t!]
\begin{center}
\caption{Speckle-interferometric measurements of resolved objects}
\label{tab1:Rastegaev_n}
\begin{tabular}{l|c|c|c|c}
\hline
Name of &  $\rho('')$ & $\Theta(\degr)$ & $\bigtriangleup m$ & Filter  \\[-4pt]
the system/ &        &      &         &      \\[-4pt]
subsystem &        &      &         &      \\
\hline
G102-20      &  0.120$\pm$0.006 & 308.0$\pm$2.8 & 3.24$\pm$0.11 & 550/20   \\
G191-55      &  0.814$\pm$0.002 & 125.1$\pm$0.3 & 2.00$\pm$0.01 & 800/100  \\
BD+19$^\circ$~1185A &  0.115$\pm$0.001 & 183.6$\pm$0.7 & 1.77$\pm$0.02 & 550/20   \\
G89-14       &  0.989$\pm$0.005 &  0.8$\pm$0.4  & 4.14$\pm$0.06 & 800/100  \\
G87-45       &  0.285$\pm$0.002 & 271.3$\pm$0.5 & 2.01$\pm$0.04 & 550/20   \\
G87-45       &  0.285$\pm$0.002 & 270.7$\pm$0.4 & 1.76$\pm$0.02 & 800/100  \\
G87-47       &  0.078$\pm$0.003 & 54.0$^{\ast}\pm$2.1 & 1.74$\pm$0.03 & 800/100  \\
G111-38AB    &  0.084$\pm$0.001 & 7.9$\pm$0.7 & 0.78$\pm$0.01 & 550/20   \\
G111-38AB    &  0.084$\pm$0.001 & 7.8$\pm$1.3 & 0.75$\pm$0.01 & 800/100  \\
G111-38AC    &  2.133$\pm$0.005 & 200.0$\pm$0.3 & 1.34$\pm$0.01 & 550/20   \\
G111-38AC    &  2.133$\pm$0.005 & 200.0$\pm$0.3 & 1.10$\pm$0.01 & 800/100  \\
G111-38BC    &  2.216$\pm$0.005 & 199.5$\pm$0.3 & 0.57$\pm$0.02 & 550/20   \\
G111-38BC    &  2.216$\pm$0.005 & 199.5$\pm$0.3 & 0.36$\pm$0.03 & 800/100  \\
G114-25      &  0.781$\pm$0.005 & 323.7$\pm$0.5 & 3.83$\pm$0.16 & 800/100  \\
\hline
\end{tabular}

\begin{tabular}{lcccc}
\multicolumn{3}{l}{$^{\ast}$ The position of the secondary component
is known with an uncertainty of  $\pm180^\circ$.}&&\\
\end{tabular}
\end{center}
\end{table*}

\begin{table*}[tbp!]
\begin{center}
\caption{Additional data on resolved stars}
\label{tab2:Rastegaev_n}
\begin{tabular}{l|c|c|c|c}
\hline
Name of the& Coordinates  & m$_{V}$& [m/H]$^{\ast}$ & Total multiplicity \\[-4pt]
system/ &   (2000.0) &      &      & of the system \\[-4pt]
subsystem &    &      &      &  \\
\hline
G102-20      & $05^h40^m09\fs7$ $+12^\circ10' 41''$ & $10.22$ & $-1.17$   & 2 \\
G191-55      & $05^h57^m28\fs6$ $+58^\circ40' 49''$ & $10.47$ & $-1.94$   & 2 \\
BD+19$^\circ$~1185A & $06^h03^m14\fs9$ $+19^\circ21' 39''$ & $9.32$  & $-1.47$   & 3 \\
G89-14       & $07^h22^m31\fs5$ $+08^\circ49' 13''$ & $10.40$ & $-1.90$   & 4 \\
G87-45       & $07^h32^m58\fs7$ $+31^\circ07' 00''$ & $11.44$ & $-1.49$   & 3 \\
G87-47       & $07^h35^m34\fs1$ $+35^\circ57' 11''$ & $10.34$ & $-1.34$   & 3 \\
G111-38      & $07^h49^m32\fs0$ $+41^\circ28' 08''$ & $8.7$   & $-1.04$   & 3 \\
G114-25      & $08^h59^m03\fs4$ $-06^\circ23' 46''$ & $11.92$ & $-2.28$   & 2\\
\hline
\multicolumn{5}{l}{$^{\ast}$  Metallicities adopted from the CLLA
catalog (\cite{clla:Rastegaev_n}).}
\end{tabular}
\end{center}
\end{table*}


\vspace*{0.3cm}
{\bf G111-38} ($07^h49^m32\fs0$ $+41^\circ28' 08''$; HIP~38195;
WDS~07495+4128). A new triple system. Hipparcos resolved it as a binary
with $\rho=2.154''$. We
resolved one of the components. The distance to the system is estimated
at $d\approx50$~pc (\cite{clla:Rastegaev_n}) or
$d\approx200$~pc (\cite{esa:Rastegaev_n}). The spectral type of the system
is G5 (\cite{simbad:Rastegaev_n}).

\vspace*{0.3cm}
{\bf G114-25} ($08^h59^m03\fs4$ $-06^\circ23' 46''$; HIP~44111).
A new binary of spectral type F7 (\cite{simbad:Rastegaev_n}). Its
heliocentric distance is $d\approx131$~pc (\cite{clla:Rastegaev_n}).

\section{MULTIPLICITY OF STARS}

\subsection{Distant Components}
For 109 of the objects considered we used additional data on the
spectroscopic multiplicity of these stars
(\cite{paperXV:Rastegaev_n}; \cite{paperXVI:Rastegaev_n}) and the data on
distant components from the WDS (\cite{wds:Rastegaev_n}) catalog.
Whereas spectroscopic and interferometric components appear to be
undoubtedly physically bound, one must treat wide visual
companions more carefully. We found a total of 43 WDS components
(in some cases several components for one object), most of which
we rejected as accidental optical projections.
Table~\ref{wds:Rastegaev_n} lists the data on the all found wide components.
 The first column gives the names of
the stars studied and the second column, all the WDS components
found. For the components found to be physically bound to the
corresponding stars columns 3 and 4 give the component separation and
magnitude difference, respectively. We adopt the latter from the
WDS catalog and they may differ slightly from the quantities
given in the corresponding references. The `$+$' and `$-$'
signs in column 5 (Status) mark the components, which we consider
to be possibly physically bound to the main star and physically
unbound optical pairs, respectively. A question mark in this
column indicates that we are not sure about our decision. The
last column  gives the references to the papers, which contain
data on the corresponding pair and confirm or disprove its
physical relationship. In all cases these are two papers
(\cite{allen:Rastegaev_n}; \cite{osorio:Rastegaev_n}) dedicated to wide
pairs of population~II stars and the Hipparcos catalog
(\cite{esa:Rastegaev_n}). Additional $\star$ symbol in this column
indicates that our observations confirm the presence of the given
component. If no references are given, it means that we made our
own decision about the physical association based on the data
provided by the  WDS catalog. To this end, we analyzed the
component magnitude difference and the change of the component
separations over the time periods covered by observations.
As a result, we left only 12 objects among the initial 43 WDS
components and used them to compute the ratios of the number of
systems of different degree of multiplicity.

\subsection{Ratio of the Systems of Different Degree of Multiplicity}

To compute the ratio of the number of systems of different degree
of multiplicity among the stars studied, we use all the data that
we gathered on the observations of these systems using different
methods. Of the 109 stars considered 24 are spectral binaries
(\cite{paperXV:Rastegaev_n}; \cite{paperXVI:Rastegaev_n}); one star
(G87-45) is a spectral triple (\cite{paperXVI:Rastegaev_n}); seven
stars are speckle-interferometric binaries, and one star
(G111-38) is a speckle-interferometric triple. Twelve stars have
companions listed in the WDS catalog. It goes without saying that
there are components detected using different methods. For
example, the  G102-20 binary with a period of 26 years
(\cite{paperXVI:Rastegaev_n}) was found both spectroscopically and
using speckle interferometry. Similarly, the outer pair in the
triple system G111-38 ($\rho\approx2''$) can be detected both
using speckle interferometry and visually.

The resulting ratio of the number of single, binary, triple, and
quadruple systems discovered using all methods among the stars of
our sample is equal to 71:28:6:1. The corresponding estimate for
F7- to G9-type disk stars uncorrected for unresolved binaries
(\cite{dm91:Rastegaev_n}) is equal to 51:40:7:2. We point out an
important difference between the two samples compared. Whereas
our sample consists of stars selected by magnitude and spatial
velocities, the sample used in \cite{dm91:Rastegaev_n} is only
distance limited (all its stars are located within  22~pc of the
Sun).

\vspace*{0.3cm}
\section{CONCLUSIONS}
We selected for observations with high angular resolution a total
of 223 high proper motion metal-poor objects  from the CLLA
catalog (\cite{clla:Rastegaev_n}). Our speckle-interferometric
observations of 109 stars made with the 6-m telescope of SAO RAS
allowed us to resolve eight stars into components and we
were the first to astrometrically resolve seven objects.
Additional data on spectral and astrometric multiplicity allowed
us to estimate the ratio of the number of single, binary, triple,
and quadruple systems to be 71:28:6:1.

In the next paper of this series we will continue to publish the
results of our speckle-interferometric observations of the star
sample presented.

\begin{acknowledgements}
This work makes use of the  SIMBAD database
and WDS catalog (\cite{wds:Rastegaev_n}).
\end{acknowledgements}

\begin{table}
\begin{center}
\caption{WDS components for stars of the sample}
\label{wds:Rastegaev_n}
\begin{tabular}{c|c|c|c|c|c}
\hline
Name & WDS companion &$\rho('')$ & $\bigtriangleup m$ & Status & References \\

G172-16      & 00386+4738OSO   7   & 8.4 & 5.85 & $+$ & \cite{osorio:Rastegaev_n} \\
G2-38        & 01270+1200LDS3282   & 24.6 & 5.5 & $+$ & \cite{allen:Rastegaev_n} \\
G172-61      & 01344+4844ES 2587   &  &  & $-$ &  \cite{osorio:Rastegaev_n} \\
G71-33       & 01452+0331LDS3306   &  &  & $-$ &  \\
G74-5        & 02104+2948BUP  29AB &  &  & $-$ &  \\
             & 02104+2948BUP  29AC &  &  & $-$ &  \\
             & 02104+2948BUP  29AD &  &  & $-$ &  \\
G37-26       & 03084+2620OSO  14   &  &  & $-$ & \cite{osorio:Rastegaev_n} \\
G246-38      & 03313+6644OSO  15   &  &  & $-$ & \cite{osorio:Rastegaev_n} \\
G95-57A/G95-57B & 03470+4126STF 443AB & 7.4 & 0.62 & $+$ & \cite{allen:Rastegaev_n} \\
             & 03470+4126STF 443AC &  &  & $-$ &  \\
             & 03470+4126FOX 135CD &  &  & $-$ & \cite{osorio:Rastegaev_n} \\
HD 25329     & 04033+3516OSO  16   &  &  & $-$ & \cite{osorio:Rastegaev_n} \\
G99-31W      & 05449+0915HDS 769   &  &  & $-$ & \cite{esa:Rastegaev_n} \\
BD+19$\degr$~1185A  & 06032+1922HDS 823Aa & 0.2 & 2.12 & $+$ & \cite{esa:Rastegaev_n}, $\star$ \\
             & 06032+1922LDS6195AB & 6.9 & 4.06 & $+$ & \cite{allen:Rastegaev_n} \\
G88-10       & 07104+2421OSO  19   &  &  & $-$ &  \cite{osorio:Rastegaev_n} \\
G89-14       & 07224+0854GIC  72AB &  &  & $-$ &  \cite{allen:Rastegaev_n} \\
             & 07224+0854ALC   2BC & 34.0 & 6.3 & $+$ & \cite{allen:Rastegaev_n} \\
G112-43/G112-44 & 07437-0004HJ 2413 & 11.8 & 1.58 & $+$ & \cite{allen:Rastegaev_n} \\
G111-38      & 07495+4128A  2468AB & 2.2 & 1.27 & $+$ & \cite{esa:Rastegaev_n}, $\star$ \\
             & 07495+4128LDS 900AB-C &  &  & $-$ &  \\
G90-25       & 07536+3036BUP 108   &  &  & $-$ &  \cite{osorio:Rastegaev_n} \\
G251-54      & 08110+7955LDS1668AB & 110.5 & 5.4 & $+$ & \cite{osorio:Rastegaev_n} \\
             & 08110+7955PWS   3AC &  &  & $-?$ &  \\
             & 08110+7955OSO  21AD &  &  & $-$ &  \cite{osorio:Rastegaev_n} \\
G40-14       & 08161+1942LDS3781   & 98.0 & 7.6 & $+$ & \cite{allen:Rastegaev_n} \\
G113-22      & 08170+0001LDS3782   &  &  & $-$ &  \\
G9-36        & 08580+2428OSO  23AB &  &  & $-$ &  \cite{osorio:Rastegaev_n} \\
             & 08580+2428OSO  23AC &  &  & $-$ &  \cite{osorio:Rastegaev_n} \\
G115-49      & 09053+3848OSO  24   &  &  & $-$ &  \cite{osorio:Rastegaev_n} \\
G120-15      & 11063+3113TDS7665AB & 7.6 & 0.68 & $+?$ &  \\
             & 11063+3113OSO  36AC &  &  & $-$ & \cite{osorio:Rastegaev_n} \\
G10-4        & 11110+0625OSO  37   &  &  & $-$ & \cite{osorio:Rastegaev_n} \\
G66-22       & 14433+0550OSO  58   & 3.2 & 3.19 & $+$ & \cite{osorio:Rastegaev_n}, $\star$ \\
G23-14       & 19518+0537OSO 112   &  &  & $-$ & \cite{osorio:Rastegaev_n} \\
G143-33      & 20084+1503OSO 117AB &  &  & $-$ & \cite{osorio:Rastegaev_n} \\
             & 20084+1503OSO 117AC &  &  & $-$ & \cite{osorio:Rastegaev_n} \\
             & 20084+1503OSO 117AD &  &  & $-$ & \cite{osorio:Rastegaev_n} \\
             & 20084+1503OSO 117AE &  &  & $-$ & \cite{osorio:Rastegaev_n} \\
             & 20084+1503LDS1033AF &  &  & $-$ &  \\
G125-64      & 20090+4252OSO 118AB &  &  & $-$ & \cite{osorio:Rastegaev_n} \\
             & 20090+4252OSO 118AC &  &  & $-$ & \cite{osorio:Rastegaev_n} \\
\hline
\end{tabular}
\end{center}
\end{table}

\onecolumn
\begin{center}
\begin{longtable}{l|c|c}
\caption{Unresolved stars} \label{unresolved:Rastegaev_n} \\
\hline \multicolumn{1}{c|}{Name} & \multicolumn{1}{c|}{Filter\,%
($\lambda/\Delta\lambda,$\,nm)} & \multicolumn{1}{c}{Epoch} \\
\hline
\endfirsthead
\caption{(Continued)} \\
 \hline
 \multicolumn{1}{c|}{Name} & \multicolumn{1}{c|}{Filter\,%
 ($\lambda/\Delta\lambda,$ nm)} & \multicolumn{1}{c}{Epoch} \\ \hline 
\endhead
\hline
\endfoot
\hline
\endlastfoot
G172-16      &  800/100 & 2006.9463 \\
G33-30       &  800/100 & 2006.9437 \\
G2-38        &  800/100 & 2006.9438 \\
G172-58      &  800/100 & 2006.9462 \\
G172-61      &  550/20; 800/100 & 2006.9462 \\
G173-10      &  550/20; 800/100 & 2006.9463 \\
G2-50        &  800/100 & 2006.9438 \\
G71-33       &  800/100 & 2006.9438 \\
G245-32      &  550/20 & 2006.9443 \\
G133-45      &  800/100 & 2006.9467 \\
G71-55       &  800/100 & 2006.9438 \\
G72-60       &  800/100 & 2006.9468 \\
G74-5        &  550/20 & 2006.9468 \\
G74-30       &  800/100 & 2006.9467 \\
G36-47       &  800/100 & 2006.9468 \\
G37-26       &  550/20 & 2006.9468 \\
G5-19        &  550/20 & 2006.9468 \\
G221-7       &  550/20 & 2006.9442 \\
G5-35        &  800/100 & 2006.9468 \\
G246-38      &  800/100 & 2006.9442 \\
G79-42       &  800/100 & 2006.9468 \\
G79-43       &  800/100 & 2006.9468 \\
G79-56       &  800/100 & 2006.9468 \\
G95-57A      &  550/20 & 2006.9414 \\
G95-57B      &  550/20 & 2006.9414 \\
HD 25329     &  550/20 & 2006.9414 \\
G82-18       &  800/100 & 2006.9415 \\
G84-29       &  550/20 & 2006.9415 \\
G191-23      &  800/100 & 2006.9448 \\
G86-39       &  800/100 & 2006.9415 \\
G86-40       &  550/20; 800/100 & 2006.9416 \\
G96-48       &  800/100 & 2006.9417 \\
G99-31W      &  550/20; 800/100 & 2006.9417 \\
G102-27      &  550/20 & 2006.9471 \\
G99-48       &  800/100 & 2006.9471 \\
G101-34      &  800/100 & 2006.9471 \\
G192-28      &  800/100 & 2006.9448 \\
G105-50      &  800/100 & 2006.9471 \\
G192-43      &  800/100 & 2006.9448 \\
G87-13       &  800/100 & 2006.9417 \\
G108-48      &  800/100 & 2006.9445 \\
G107-50      &  800/100 & 2006.9475 \\
G87-27       &  550/20; 800/100 & 2006.9472 \\
G88-10       &  800/100 & 2006.9472 \\
G108-58      &  800/100 & 2006.9445 \\
G88-27       &  800/100 & 2006.9472 \\
G90-3        &  800/100 & 2006.9472 \\
G88-32       &  800/100 & 2006.9418 \\
BD-1$^\circ$~1792    &  800/100 & 2006.9446 \\
G112-43      &  800/100 & 2006.9446 \\
G112-44      &  800/100 & 2006.9446 \\
G90-25       &  550/20; 800/100 & 2006.9473 \\
G251-54      &  800/100 & 2006.9475 \\
G234-24      &  800/100 & 2006.9475 \\
G40-8        &  800/100 & 2006.9446 \\
G234-28      &  800/100 & 2006.9448 \\
G40-14       &  800/100 & 2006.9446 \\
G113-22      &  800/100 & 2006.9446 \\
G194-22      &  800/100 & 2006.9448 \\
BD+25$^\circ$~1981   &  550/20; 800/100 & 2006.9473 \\
G46-5        &  800/100 & 2006.9474 \\
G115-34      &  800/100 & 2006.9473 \\
G9-36        &  800/100 & 2006.9473 \\
G114-26      &  550/20 & 2006.942 \\
G115-49      &  800/100 & 2006.9473 \\
G46-31       &  800/100 & 2006.9476 \\
G41-41       &  800/100 & 2006.9476 \\
G195-34      &  800/100 & 2006.9422 \\
G48-29       &  800/100 & 2006.9476 \\
G116-45      &  800/100 & 2006.9473 \\
G161-73      &  800/100 & 2006.9476 \\
G43-3        &  545/30 & 2006.2759 \\
G53-41       &  800/100 & 2006.9421 \\
G44-30       &  800/100 & 2006.9422 \\
G58-23       &  800/100 & 2006.9422 \\
G196-48      &  800/100 & 2006.9422 \\
G58-25       &  550/20 & 2006.9422 \\
G146-76      &  800/100 & 2006.9477 \\
G253-41      &  550/20; 800/100 & 2006.9422 \\
G120-15      &  800/100 & 2006.9449 \\
G10-4        &  800/100 & 2006.9449 \\
BD+36$^\circ$~2165   &  800/100 & 2006.9449 \\
HD 97916*    &  550/20 & 2006.945 \\
G56-30       &  800/100 & 2006.9449 \\
G254-24      &  800/100 & 2006.9423 \\
G147-62      &  800/100 & 2006.945 \\
G121-12      &  800/100 & 2006.9477 \\
G176-53      &  800/100 & 2006.9423 \\
G122-51      &  550/20 & 2006.945 \\
G66-22       &  545/30; 800/110 & 2006.3747 \\
G166-45      &  545/30 & 2006.3747 \\
G16-13       &  545/30 & 2006.3748 \\
G16-20       &  545/30 & 2006.3749 \\
G153-21      &  545/30 & 2006.3749 \\
G170-47      &  545/30 & 2006.4488 \\
G23-14       &  800/110 & 2006.4517 \\
G23-20       &  800/110 & 2006.4517 \\
G24-3        &  800/110 & 2006.4518 \\
G143-33      &  800/110 & 2006.4518 \\
G125-64      &  800/110 & 2006.4518 \\
G171-15      &  800/100 & 2006.9463 \\
\end{longtable}
\end{center}
\clearpage
\twocolumn

\end{document}